\documentclass[12pt]{iopart}
\usepackage{times}
\usepackage{cite}
\usepackage{graphicx,psfrag}

\usepackage{bbm}

\begin{document}
\title{Transmission Through a Quantum Dynamical Delta Barrier}
\author{T. Brandes and J. Robinson}
\address{Dept. of Physics, UMIST, PO Box 88, Manchester M60 1QD, UK}

\begin{abstract}
We discuss electron scattering in 
a one-dimensional delta barrier potential with either
time-dependent coupling constant (classical model) or  
a coupling constant that is linear in a boson coordinate (quantum model). 
We find an exact continued fraction solution and Fano like
resonances in the transmission coefficient. In the quantum model, energies for  perfect
transmission exist below the first sideband threshold. 
\end{abstract}

\section{Introduction and Model}
Simple models for the interaction between fermions and bosons continue to  
be fascinating, as often very non-trivial results can be obtained from even the
most primitive Hamiltonians. 
Exactly solvable models for the interaction of photons or phonons with electrons 
in quantum dots \cite{SZC00} or quasi-one-dimensional systems
\cite{GSL89,Bag90} provide the best examples, as they often 
provide a deep insight into rich and complex physics.

In this contribution, we re-consider a simple model for 
a single electron of mass $m$ in one dimension that 
interacts with a delta-barrier through a coupling parameter
that itself is a dynamical quantity.
The  Hamiltonian is written as
\begin{eqnarray}\label{Hamiltonian}
  H=\frac{p^2}{2m}+\delta(x)\left\{ g_0+g_1[a^{\dagger}+a]\right\}+\Omega a^{\dagger}a.
\end{eqnarray}
Here, $a^{\dagger}$ creates a boson of frequency $\Omega$ 
and $g_1[a^{\dagger}+a]$ is a dynamical contribution on  top of the static
coupling constant $g_0$. The constant zero point energy is omitted since it merely 
shifts the energy scale by $\Omega/2$.
The lattice version of this model was  originally introduced
by Gelfand, Schmitt-Rink and Levi \cite{GSL89} years ago in the study of tunneling in presence of phonons,
and was shortly afterwards compared to 
a corresponding time-dependent classical Hamiltonian \cite{LTJ90}, the continuous version of 
which reads
\begin{eqnarray}\label{Hamiltonianc}
  H_{\rm cl}(t)=\frac{p^2}{2m}+\delta(x)\left\{ g_0+2g_1\cos(\Omega t)
\right\}.
\end{eqnarray}
$H_{\rm cl}(t)$
is obtained as the interaction picture Hamiltonian of Eq.(\ref{Hamiltonian}) 
with respect to $H_B=\Omega a^{\dagger}a$,
after replacing the boson operators by $a^{\dagger}=a=1$.

In its 
time-dependent version, Eq.(\ref{Hamiltonianc}) has subsequently been used as a model for
scattering in quasi-one-dimensional quantum wires by Bagwell and co-workers \cite{Bag90,BL92},
who found Fano-type resonances in the transmission coefficient as a function of 
the energy of an incident electron. It soon turned out that the scattering properties 
of this Hamiltonian are  quite intriguing as they very much depend on the relative sign
and strength of the two coupling parameters $g_0$ and $g_1$. 
The interplay between evanescent modes,  quasibound states \cite{BL92},
and the behaviour of the transmission amplitude in the complex energy plane
\cite{MR01,KPSS02} have been studied recently. 

Our focus here is on the quantum version Eq. (\ref{Hamiltonian}) of the model 
and its peculiarities in comparison with $H_{\rm cl}$. It turns out that
beside transmission zeroes, there are points of perfect transparency in the 
Fano resonance that only appear in the model $H$ but not in $H_{\rm cl}$. 
Perfect transmission and Fano resonances  have been found
recently in the transmission of phonons through non-linear chains
without delta impurities \cite{KBM97,KK01}. Although not discussed in detail here,
these results indicate that there still is rich and non-trivial behavior 
to be discovered from models like Eq.(\ref{Hamiltonian}).

\section{Transmission Coefficient}
The total wave function $|\Psi\rangle$ of the coupled electron-boson system can be 
expanded in the oscillator basis $\{ |n\rangle \}$ as
\begin{eqnarray}\label{totalwave}
  \langle x|\Psi\rangle=\sum_{n=0}^{\infty}\psi_n(x) |n\rangle
\end{eqnarray}
with wave function coefficients $\psi_n(x)$ depending on the position $x$ of the electron.
We solve the stationary Schr\"odinger equation at total energy $E>0$, 
implying a scattering condition
for the electron part of the wave function in demanding that there is no electron incident from the right.
For $x\ne 0$, the $\psi_n(x)$ are superpositions of plane waves if $E$ is above the threshold
for the $n-th$ boson energy, 
\begin{eqnarray}\label{waveprop}
 \psi_n(x<0)&=&a_ne^{ik_nx}+b_ne^{-ik_nx}\nonumber\\
 \psi_n(x>0)&=&t_ne^{ik_nx},\quad k_n\equiv \sqrt{E-n\Omega},\quad
E>n\Omega,
\end{eqnarray}
whereas normalizabale evanescent modes occur if $E$ is below the threshold,
\begin{eqnarray}\label{waveevan}
 \psi_n(x<0)&=&b_ne^{\kappa_nx}\nonumber\\
 \psi_n(x>0)&=&t_ne^{-\kappa_nx},\quad \kappa_n\equiv \sqrt{n\Omega-E},\quad
E<n\Omega.
\end{eqnarray}
Here and in the following we set $\hbar = 2m =1$.
We impose the condition that the boson is in its ground state for an electron incoming from the left,
\begin{eqnarray}
  a_n=\delta_{n,0},
\end{eqnarray}
where we set the corresponding amplitude $A=A_0$ to unity. 
Continuity of $\psi_n(x)$ at $x=0$ yields $a_n+b_n=t_n$ for all $n$,
whereas the jump in derivative of $\psi_n(x)$ across the delta barrier
leads to a recursion relation for the transmission amplitudes $t_n$,
\begin{eqnarray}\label{recursion}
  g_1\sqrt{n}t_{n-1}+(g_0 -2i\gamma_n)t_n + g_1 \sqrt{n+1} t_{n+1}&=&-2i\gamma_n \delta_{n,0}
\end{eqnarray}
where the $\gamma_n$ are real (imaginary) above (below) the boson energy $n\Omega$,
\begin{eqnarray}
\gamma_n&=& k_n \theta(E-n\Omega)+i\kappa_n\theta(n\Omega-E).  
\end{eqnarray}
The total transmission coefficient $T(E)$ is obtained from the sum over all {\em propagating} modes,
\begin{eqnarray}\label{transmission}
  T(E) = \sum_{n=0}^{[E/\Omega]}\frac{k_n(E)}{k_0(E)}|t_n(E)|^2,
\end{eqnarray}
where the sum runs up to the largest $n$ such that $k_n$ remains real.

\section{Matrix Representation and Continued Fractions}
Although Eq.(\ref{transmission}) is a finite sum, its evaluation requires the solution of the 
{\em infinite} recursion relation Eq.(\ref{recursion}) due to the fact  that the propagating modes
are coupled to all evanescent modes. The transmission amplitudes can be determined from the linear 
equation 
\begin{eqnarray}\label{matrix}
M{\bf t} &=& {\bf a},\quad {\bf t}=(t_0,t_1,t_2,...),\quad
{\bf a}= (-2i\gamma_0,0,0,...)\nonumber\\
M&=&\begin{array}{r}
    \left(
    \begin{array}{ccccc}
        g_0-2i\gamma_0 & \sqrt{1}g_1 & 0 &    \\
        \sqrt{1}g_1 & g_0-2i\gamma_1 & \sqrt{2}g_1 & 0   \\
        0 & \sqrt{2}g_1 & g_0-2i\gamma_2 & \ddots   \\
          & 0  & \ddots & \ddots 
    \end{array}\right).
\end{array}
\end{eqnarray}
Numercally, this can easily been solved by truncation of the matrix $M$.
Alternatively, one can solve Eq.(\ref{matrix}) recursively which actually is
numerically more efficient. For example, 
the result for the zero-channel transmission amplitude $t_0(E)$
can be written in a very intuitive form:  defining the `Greens function' $G_0(E)$ by
\begin{eqnarray}\label{G0def}
  G_0(E) \equiv [-2i\gamma_0(E) +g_0]^{-1},
\end{eqnarray}
one can  write $t_0(E)$ with the help of a recursively defined 
`self energy' $\Sigma^{(N)}(E)$,
\begin{eqnarray}\label{selfenergy}
  t_0(E) &=& \frac{-2i\gamma_0(E)}{G^{-1}_0(E)-\Sigma^{(1)}(E)},
\quad  \Sigma^{(N)}(E) = \frac{Ng_1^2}{G^{-1}_0(E-N\Omega)- \Sigma^{(N+1)}(E)}.
\end{eqnarray}
In fact, using $\gamma_n(E)=\gamma_0(E-n\Omega)$,
the self energy $\Sigma^{(1)}(E)$ can be represented as a continued fraction 
\begin{eqnarray}\label{sigmacontinued}
  \Sigma^{(1)}(E) &=&
\frac{g_1^2}{\displaystyle G^{-1}_0(E-\Omega) - \frac{2g_1^2}{\displaystyle G^{-1}_0(E-2\Omega) -
\frac{3g_1^2}{\displaystyle G^{-1}_0(E-3\Omega) - \frac{4g_1^2}{\displaystyle \ddots}}}}\,.
\end{eqnarray}
This demonstrates that $t_0(E)$ depends on $g_1$ only through $g_1^2$.

Truncating the matrix $M$ to a $N\times N$ matrix corresponds to the approximation that
sets $\Sigma^{(N)}(E)\equiv 0$ and recursively solves Eq. (\ref{selfenergy}) for
$\Sigma^{(N-1)}(E)$ down to $\Sigma^{(1)}(E)$. For example, truncating at $N=2$ we obtain the
approximation
\begin{eqnarray}\label{t0approx}
  t_{0,{N=2}}(E) &=& \frac{-2i\gamma_0(E)}{G^{-1}_0(E)-\Sigma^{(1)}_{N=2}(E)}
=\frac{-2i\gamma_0(E)}{-2i\gamma_0(E)+g_0 -\frac{\displaystyle g_1^2}{\displaystyle -2i\gamma_1(E)+g_0}}.
\end{eqnarray}
An important observation can be made with respect to the stability of our theory for large
coupling constants $g_1$. In fact, the truncation at $N+1$ 
is only consistent if the truncated self energy $\Sigma^{(N)}(E)$ is a small correction to
the inverse `free propagator',
\begin{eqnarray}
Ng_1^2/|G^{-1}_0(E-N\Omega)|<  |G^{-1}_0(E-(N-1)\Omega)|,
\end{eqnarray}
which by use of Eq. (\ref{G0def}) at large $N$ implies $Ng_1^2<4N\Omega$ or $g_1<2\sqrt{\Omega}$. 
The tridiagonal form of the matrix, Eq. (\ref{matrix}), actually implies that the method
based on the recursion Eq. (\ref{recursion}) is perturbative in the coupling $g_1$ to the boson.  
We conjecture that for $g_1$ above the critical value, the perturbation based 
on the oscillator basis $\{ |n\rangle \}$  used here breaks down.
A similar breakdown of numerical approaches that start from
a weak coupling regime in single boson Hamiltonians is known from the
Rabi Hamiltonian \cite{rabi1}, i.e. the coupling of a single boson mode to a spin $1/2$.

We mention that the lattice version of the present model would be a natural starting point for a more detailed
analysis of the strong coupling (small polaron) limit.

\section{Comparison to the Classical Case}
\begin{figure}[t]
\includegraphics[width=0.5\textwidth]{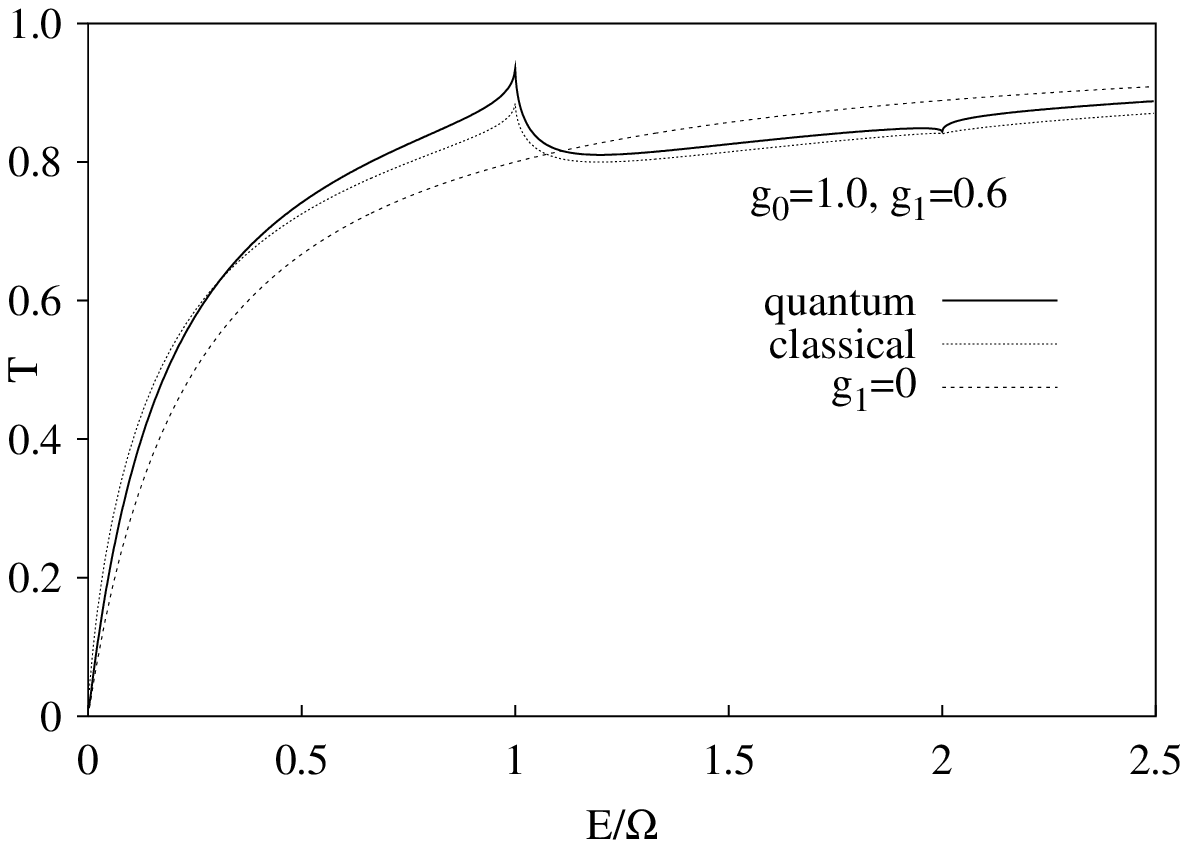}
\includegraphics[width=0.5\textwidth]{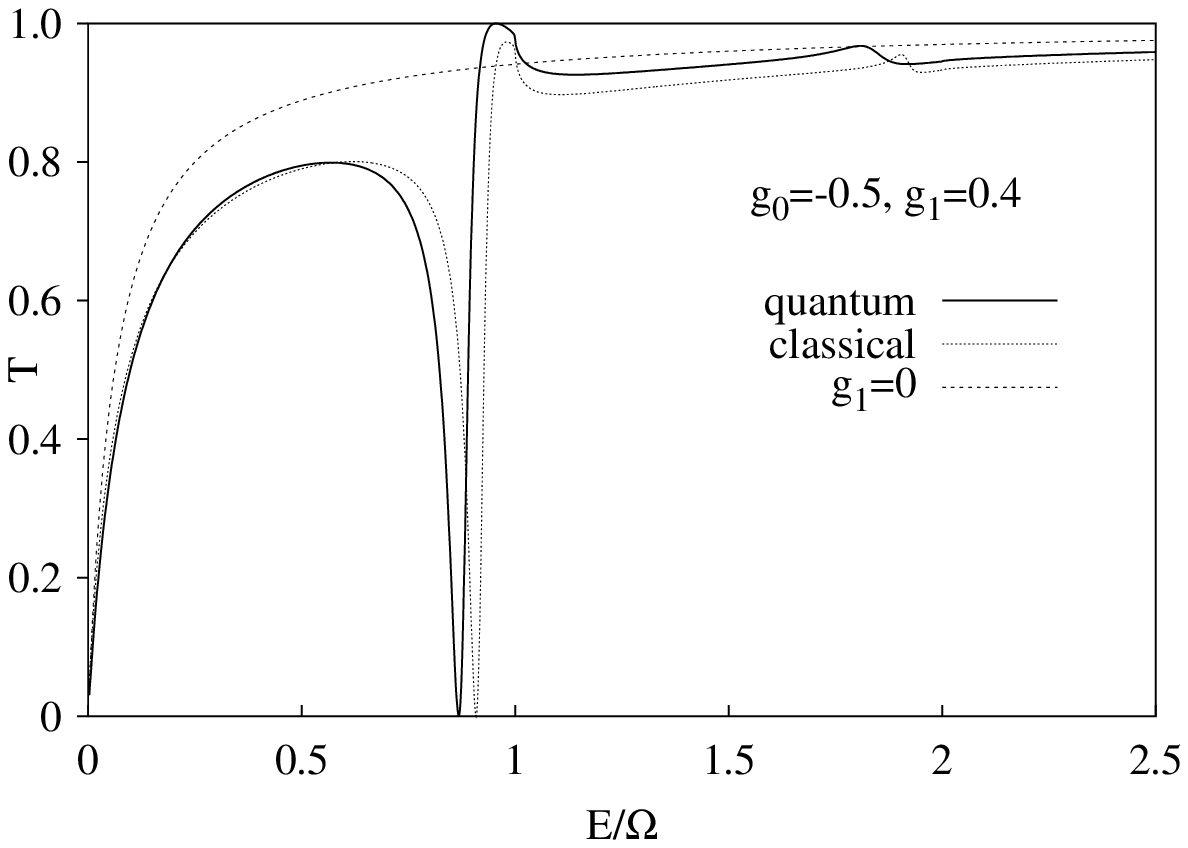}
\caption[]{\label{delta1.eps}Transmission coefficient through a dynamical 
one-dimensional delta barrier
with repulsive ($g_0>0$, left) and attractive ($g_0<0$, right) static part, cf. Eq. (\ref{Hamiltonian}) and
(\ref{Hamiltonianc}). 
$E$ is the energy of the incident particle.}
\end{figure} 
The recursion relation corresponding to Eq. (\ref{recursion}) for the classical time-dependent
Hamiltonian, Eq. (\ref{Hamiltonianc}), was derived and discussed by Bagwell and Lake \cite{BL92},
\begin{eqnarray}
  g_1 t_{n-1}+(g_0 -2i\gamma_n)t_n + g_1 t_{n+1}&=&-2i\gamma_n \delta_{n,0},\quad n=0,\pm 1,\pm 2,...
\end{eqnarray}
Here, $t_n$ is the coefficient of the time-dependent electron wave function in photon side-band $n$, 
where $n$ runs through positive {\em and negative} integers $n$. In further contrast to
the recursion relation Eq. (\ref{recursion}), there are no factors 
 $\sqrt{n}$ and $\sqrt{n+1}$ multiplying the coupling constant $g_1$. This latter fact is 
an important difference to the quantum case where these terms lead to the 
factors $N$ that multiply $g_1^2$ in the self energies $\Sigma^{(N)}(E)$, Eq. (\ref{selfenergy}).
This difference is eventually responsible for the breakdown of the perturbative approach 
for large $g_1$ in the quantum case. 
 
A continued fraction representation of $t_0(E)$ for the classical case
has been derived recently by Martinez and Reichl \cite{MR01}. The corresponding matrix 
defining the transmission amplitudes ${\bf t}_{\rm cl}=(...,t_{-2},t_{-1},t_0,t_1,t_2,...)$
in the classical case is the infinite tridiagonal matrix $M_{\rm cl}$ with $ g_0-i\gamma_{n}$
on the diagonal and $g_1$ on the lower and upper diagonals,
\begin{eqnarray}\label{matrixcl}
M_{\rm cl}&=&\begin{array}{r}
    \left(
    \begin{array}{ccccc}
\ddots  & \ddots & 0 & &\\
      \ddots &  g_0-2i\gamma_{-1} & g_1 & 0 &    \\
   0  &   g_1 & g_0-2i\gamma_0 & g_1 & 0   \\
     &   0 & g_1 & g_0-2i\gamma_1 & \ddots   \\
      &    & 0  & \ddots & \ddots 
    \end{array}\right).
\end{array}
\end{eqnarray}
Fig. (\ref{delta1.eps}) shows a comparison between the transmission
coefficient $T(E)$, Eq. (\ref{transmission}), for the quantum  and the classical barrier.
In the repulsive case with $0<g_1<g_0$, the dynamical part of the barrier  is essentially a
weak perturbation to the unperturbed $(g_1=0)$ case. Additional structures (cusps) appear
at the boson (photo side-band) energies $n\Omega$ although the overall $T(E)$-curve resembles the
$(g_1=0)$ case.

The more interesting case occurs for barriers with an attractive static part, $g_0<0$ 
(Fig. (\ref{delta1.eps}), right). A Fano type resonance appears below the first 
threshold $E=\Omega$ where the transmission coefficient has a zero in both the classical
and the quantum case. In the classical case, 
this is a well-known phenomenon \cite{BL92}: the transmission zero 
for weak coupling (small $g_1$)
shows up when the Fano resonance condition
\begin{eqnarray}\label{Fano1}
 2\kappa_1(E)+g_0=0 
\end{eqnarray}
is fulfilled.
There, the energy of the electron in the first side channel ($n=1$) 
coincides with the bound state of the attractive delta barrier potential,
$E-\Omega = -g_0^2/4$.
In the quantum case, the self energy in Eq.(\ref{selfenergy}) diverges at the zeros
of $T(E)$, 
\begin{eqnarray}\label{Fano1a}
  [{\Sigma^{(1)}(E)}]^{-1}=0.
\end{eqnarray}
For $g_1\to 0$, ${\Sigma^{(1)}}(E)\to \Sigma^{(1)}_{N=2}(E)= g_1^2/(2\kappa_1(E)+g_0)$, cf. 
Eq.(\ref{t0approx}), and the two conditions
Eq.(\ref{Fano1}) and Eq.(\ref{Fano1a}) coincide.

\section{Perfect Transparency}
\begin{figure}[t]
\centering
\includegraphics[width=0.45\textwidth]{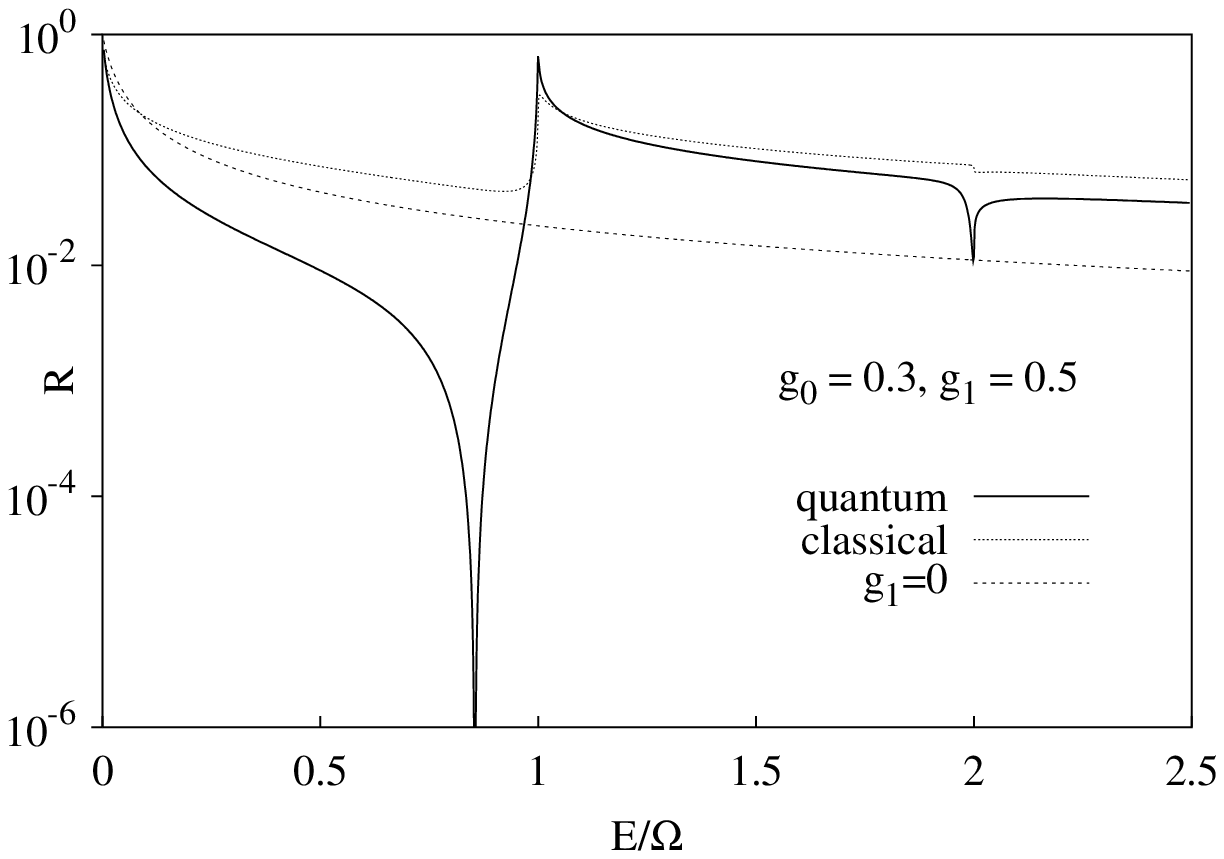}
\includegraphics[width=0.45\textwidth]{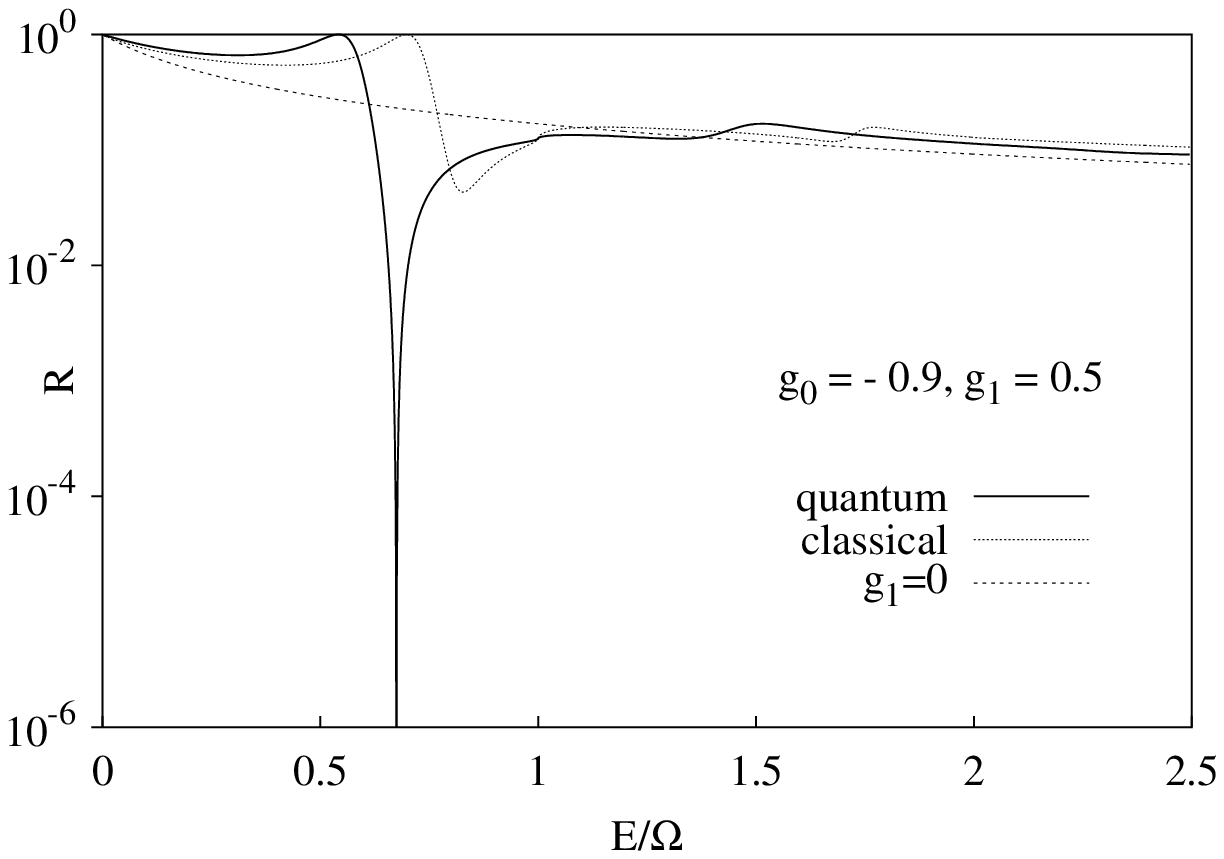}
\caption[]{\label{delta3.eps}Logarithmic plot of reflection coefficient $R\equiv 1-T$ for dynamical delta
barrier with static repulsive ($g_0>0$, left) and attractive ($g_0<0$, right) core. }
\end{figure}

\begin{figure}[t]
\psfrag{E}{\hspace*{-8mm}$E/\Omega$}
\psfrag{g}{$g_0$}
\psfrag{20}{$20$}
\psfrag{1}{\hspace*{-1mm}$1$}
\psfrag{0.5}{$0.5$}
\psfrag{-1}{\hspace*{-3mm}$-1$}
\psfrag{0}{\hspace*{-1mm}$0$}
\includegraphics[width=0.95\textwidth]{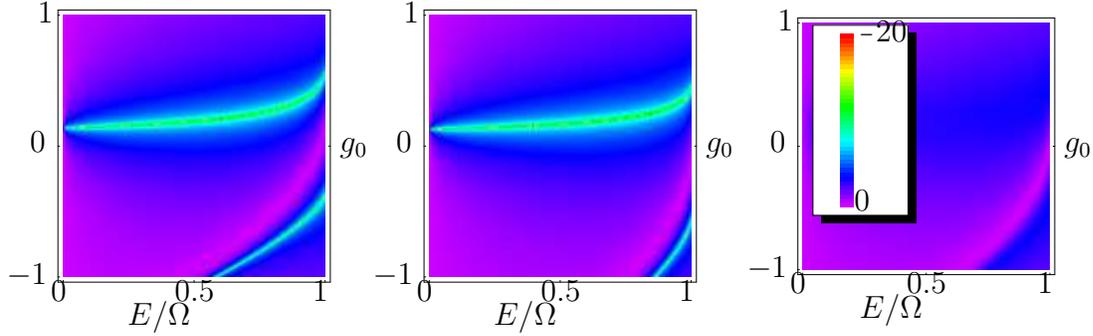}
\caption[]{\label{delta4.eps}Density plot of $\ln R$ (reflection coefficient) for the 
quantum delta barrier at $g_1=0.5$.  Exact solution from Eq. (\ref{selfenergy}) (left),
from the $N=2$ truncation Eq. (\ref{t0approx}) (center), and from the classical model Eq.(\ref{Hamiltonianc})
(right).
The light `ridges' correspond to curves of perfect transmission $T$, cf. Eqs. (\ref{perfecta}) and 
Eqs. (\ref{perfectb}).}
\end{figure}

The most interesting feature in the scattering properties of the dynamical quantum barrier 
is the appearance of an energy  close to the 
first channel $(n=1)$ threshold where perfect transmission $T(E)=1$ occurs.
This is clearly visible in the vanishing of the reflection coefficient,
$R(E)\equiv 1-T(E)$, in the logarithmic plot Fig. (\ref{delta3.eps}). 
For a repulsive static part, $g_0=0.3$, this occurs at an energy below 
the energy where the reflection
coefficient comes close to unity, and above that energy if the static part is attractive
($g_0=-0.9$).
In contrast, in the classical case the reflection coefficient never reaches 
zero in neither the repulsive nor the attractive case.
This contrast  becomes even more obvious in a two-dimensional plot 
where the zeros in $R$ correspond to `ridges'
in the $g_0$-$E$ plane, cf. Fig. (\ref{delta4.eps}).

Perfect transparency ($R=1-T=0$) can be understood by considering the transmission amplitude
$t_0(E)$ which determines the total transmission below the first sideband
threshold. Recalling that $t_0(E)=-2ik_0/(-2ik_0+g_0-\Sigma^{(1)}(E))$, 
in the quantum case the transmission coefficient becomes unity when
\begin{eqnarray}\label{perfect}
  g_0-\Sigma^{(1)}(E)=0.
\end{eqnarray}
Our exact continued fraction expression for the self energy, Eq.(\ref{sigmacontinued}), 
implies that for $0<E<\Omega$, 
$\Sigma^{(1)}(E)$ is real because $G_0^{-1}(E-n\Omega)=2\sqrt{n\Omega-E}+g_0$ is real
for $n\ge 1$. The condition Eq.(\ref{perfect}) then means that the self energy
exactly renormalizes the static part $g_0$ of 
the scattering potential to zero.
  
For small $g_1$, we can use
our perturbative expression corresponding to truncating the matrix $M$, Eq.(\ref{matrix}), to a
two-by-two matrix. The perfect transparency condition Eq.(\ref{perfect}) then becomes
\begin{eqnarray}
  \label{perfecta}
  g_0 - \frac{g_1^2}{2\kappa_1(E)+g_0}=0,\quad 0<E<\Omega,\quad (N=2 \mbox{ truncation.}),
\end{eqnarray}
which determines the position of the perfect transmission energy.
The solution of Eq.(\ref{perfecta}) defines two curves in the
$E$--$g_0$-plane with perfect transmission for $0<E<\Omega$,
\begin{eqnarray}\label{perfectb}
  g_0=-\sqrt{\Omega-E}\pm \sqrt{\Omega-E+g_1^2}.
\end{eqnarray}
These two curves can be clearly identified in the logarithmic density plots
of the reflection coefficient $R=1-T$, cf. Fig. (\ref{delta4.eps}). The $N=2$ approximation to the 
transmission amplitude, Eq.(\ref{t0approx}), turns out to 
reproduce these features quite well even at moderate coupling constants
$g_1$.

\section{Conclusions}
The above analysis of the two models
Eq. (\ref{Hamiltonian}) and  Eq. (\ref{Hamiltonianc}) has revealed some interesting 
differences between scattering properties of simple electron-boson models and their classical counter-part.
The strong coupling limit of the quantum model and its extension to 
more complicated situations like multi-channel scattering remain to be explored.

This work was supported by the UK project EPSRC GR44690/01, and the 
UK Quantum Circuits Network.


\end{document}